\documentclass[reprint,amsmath,amssymb,aps]{revtex4-2}

\usepackage{graphicx}% Include figure files
\usepackage{dcolumn}% eqnarray table columns on decimal point
\usepackage{bm}% bold math
\usepackage{float}% for [H]
\usepackage{xcolor}% for color        {\color{blue} A}
\usepackage{tikz}
\usepackage[colorlinks = true,
linkcolor = blue,
urlcolor  = blue,
citecolor = blue,
anchorcolor = blue]{hyperref}

\begin{document}

\title{Quadratic pair-breaking absorption edge from Anderson localization in a superconducting wire}

\author{Bahruz Suleymanli}
\email{bahruz.suleymanli@gmail.com}
\affiliation{Department of Physics, Yıldız Technical University, 34220 Esenler, Istanbul, Türkiye}

\author{B. Tanatar}
\email[Corresponding author: ]{tanatar@fen.bilkent.edu.tr}
\affiliation{Department of Physics, Bilkent University, 06800 Ankara, Türkiye}

\date{\today}

\begin{abstract}
We show that Anderson localization fundamentally reshapes dissipative
superconducting electrodynamics, replacing the linear Mattis--Bardeen
pair-breaking onset with a parametrically weaker quadratic Mott edge while
leaving the quasiparticle spectrum unchanged. We derive this behavior for a weakly disordered single-channel wire with a
spatially uniform $s$-wave pair potential using a Nambu-space extension of
the Berezinskii diagram technique that resums elastic impurity scattering
to all orders. Conservation of the Bogoliubov branch collapses the Nambu diagram hierarchy
onto an exactly solvable localization problem. The localization length
remains equal to its normal-state value, whereas the localization time
diverges at the gap edge. The new absorption edge results from the product of the superconducting
pair-creation coherence factor and the Mott-suppressed current matrix
elements between localized orbitals, which generates the characteristic
double logarithm. Localization also reduces the Hebel--Slichter coherence
peak. The resulting theory provides an all-orders benchmark for microwave
and spin-relaxation experiments on localized superconducting nanowires.
\end{abstract}

\maketitle

%%%%%%%%%%%%%%%%%%%%%%%%%%%%%%%%%%%%%%%%%%%%%%%%%%%%%%%%%%%%%%%%%%%%%%%
Recent experiments have established semiconductor--superconductor
nanowires as a controlled platform for one-dimensional superconductivity.
Epitaxial growth and interface engineering produce hard induced gaps
\cite{Krogstrup2015,Chang2015,Gul2017}, electrostatic gates restrict
transport to one or a few transverse modes
\cite{Zuo2017,Heedt2021}, and microwave spectroscopy resolves individual
and pair transitions between Bogoliubov excitations
\cite{vanWoerkom2017,Hays2018,Zellekens2022,MatuteCanadas2022}.
Finite-frequency measurements also reveal a sharp contrast between
spectral and dissipative properties. Conventional dirty superconductors
can follow Mattis--Bardeen electrodynamics across the excitation gap
\cite{Thiemann2018}, whereas disordered TiN, NbN, MoN, and
InO$_x$ films exhibit anomalous dissipative spectral weight,
spatially inhomogeneous response, and disorder-enabled collective modes
beyond a uniform Mattis--Bardeen description
\cite{Driessen2012,Coumou2013,Sherman2015,Simmendinger2016,
Cheng2016,Katsumi2024}.
Recent experiments further show that localized quasiparticles can control
relaxation and microwave loss
\cite{Rooij2025,Larson206}.

On the theoretical side, symmetry-class approaches establish
localization of quasiparticles in quasi-one-dimensional superconductors
\cite{Altland1997,Brouwer2000}, and localized superconductors can exhibit
nonstandard low-frequency ac conductivity
\cite{Schmitz2018}.  These results do not determine how localization acts
at the pair-breaking threshold of a hard-gapped wire.  This raises the
fundamental question of whether Anderson localization can reorganize
dissipative superconducting electrodynamics while leaving the excitation
spectrum intact.  The answer is also relevant to superconducting quantum
circuits, where nonequilibrium quasiparticles generated by pair-breaking
radiation and phonons produce energy relaxation, parity switching, and
microwave loss
\cite{Martinis2009,Serniak2018,Patel2017,Pan2022,Liu2024}.
We resolve this problem in the analytically controlled regime of a weakly
disordered single-channel superconducting wire.
%%%%%%%%%%%%%%%%%%%%%%%%%%%%%%%%%%%%%%%%%%%%%%%%%%%%%%%%%%%%%%%%%%%%%%%

%%%%%%%%%%%%%%%%%%%%%%%%%%%%%%%%%%%%%%%%%%%%%%%%%%%%%%%%%%%%%%%%%%%%%%%
Anderson showed that quantum interference can arrest diffusion in a
random medium \cite{Anderson1958}, while in one dimension arbitrarily weak
disorder localizes all single-particle states
\cite{MottTwose1961}.  To retain these interference processes in the
superconducting problem, we extend the Berezinskii diagram technique to
Bogoliubov quasiparticles.  The method orders impurity vertices along the
one-dimensional coordinate and resums the resulting retarded--advanced
diagram blocks to all orders in elastic scattering
\cite{Berezinskii1974}. Its arbitrary-frequency formulation yields both
the complete four-point function of a weakly disordered normal conductor
and the full evolution of a wave packet into a localized state
\cite{Nakhmedov1987}.  We show that this all-orders resummation remains
exactly reducible despite the two-branch Nambu structure of a
superconductor.
%%%%%%%%%%%%%%%%%%%%%%%%%%%%%%%%%%%%%%%%%%%%%%%%%%%%%%%%%%%%%%%%%%%%%%%

%%%%%%%%%%%%%%%%%%%%%%%%%%%%%%%%%%%%%%%%%%%%%%%%%%%%%%%%%%%%%%%%%%%%%%%
We consider a single-channel wire with a uniform real gap and scalar
potential disorder,
\begin{equation}
 \hat H_{\rm BdG}
 =\left[\frac{\hat p^2}{2m}-\mu+U(x)\right]\tau_3
 +\Delta\tau_1 .
 \label{eq:bdg}
\end{equation}
Forward and backward scattering define the normal-state mean free paths
$l_1$ and $l_2$, with $k_Fl_{1,2}\gg1$. Starting from the exact
coordinate-space Gor'kov functions, we linearize the propagator near the
Fermi points.  For $\omega>\Delta$, it becomes
\begin{align}
 \hat G^{\pm}(x|\omega)
 &\simeq\mp\frac{i}{v_g}e^{\pm i\eta|x|/v_F}
 \sum_{s=\pm}e^{\pm is k_F|x|}\hat M_s,
 \label{eq:linearized-propagator}
\end{align}
where
$
 \hat M_s=
 \left(1+s\eta\tau_3/\omega+\Delta\tau_1/\omega\right)/2,
 \qquad
 \eta=\sqrt{\omega^2-\Delta^2},
 \qquad
 v_g=v_F\eta/\omega.
$
The projectors $\hat M_s^2=\hat M_s$ select the two solutions
$\xi=s\eta$ of the Bogoliubov dispersion.  The label $s$ distinguishes
the particle-like and hole-like branches.

Eq.~\eqref{eq:linearized-propagator} is the building block of the
disorder expansion for the retarded--advanced correlator
$\langle\hat G^-\otimes\hat G^+\rangle$.  At a given order, the two Green
functions form chains of propagator segments separated by potential
vertices $U(x)\tau_3$.  Gaussian disorder averaging pairs these vertices.
Because the system is one-dimensional, their coordinates can be ordered
along the wire.  Cutting an ordered diagram at a fixed coordinate
classifies it by the number $m$ of retarded--advanced propagator pairs
crossing the cut.  Summing all diagrams with the same $m$ produces a
closed recursion that resums the four-point function to all orders
\cite{Berezinskii1974,Nakhmedov1987}.  The new ingredient in these blocks
is the Nambu matrix carried by each propagator segment.

Consider a segment that reaches an impurity vertex on branch $s$ and leaves
on branch $s'$.  The projectors on the two sides of the vertex give the
Nambu factor $\hat M_s\tau_3\hat M_{s'}$.  On the Bogoliubov mass shell,
\begin{equation}
 \hat M_s\tau_3\hat M_{s'}
 =\delta_{ss'}\,s\frac{\eta}{\omega}\hat M_s .
 \label{eq:branch-conservation}
\end{equation}
The factor $\delta_{ss'}$ shows that scalar potential disorder cannot
convert a particle-like branch into a hole-like branch.  The branch label
is therefore conserved along every Green-function chain and at every
order in the disorder expansion.  For $s=s'$, each vertex contributes
$s\eta/\omega$.  A disorder contraction contains two vertices and hence
the factor $(\eta/\omega)^2$.  The two propagator amplitudes associated
with that contraction contain
$1/v_g=(\omega/\eta)/v_F$ and supply the inverse factor
$(\omega/\eta)^2$.  Their product is exactly unity.  After projection onto
a conserved branch, every spatially ordered diagram therefore retains its
normal-state scalar weight.

The forward- and backward-scattering lengths are consequently unchanged,
$l_{1,2}^{\rm sc}=l_{1,2}$, and every quasiparticle localizes on the
normal-state length
$
 L_{\rm loc}=2l_2.
$
Superconductivity changes the propagation phases instead.  A
quasiparticle of energy $\varepsilon$ accumulates the slow phase
$\eta(\varepsilon)x/v_F$, so Green functions separated in energy by
$\nu$ acquire the mismatch
$
 \eta(\varepsilon+\nu)-\eta(\varepsilon)
 \simeq\omega\nu/\eta.
$
The equal-branch contribution therefore contains the effective
localization time
$
 \widetilde{\tau}_2=\tau_2\omega/\eta.
$

Branch conservation separates
$\langle\hat G^-\otimes\hat G^+\rangle$ into two equal-branch sectors
controlled by
$s_{\parallel}=2i\widetilde{\tau}_2\nu$ and two mixed-branch sectors
controlled at small $\nu$ by
$
 s_{\perp}\simeq4i\eta\tau_2,
 \qquad
 k_+-k_-=2\eta/v_F.
$
Each sector reduces to the scalar recursion introduced above.
Superconductivity enters through $s_{\parallel}$, $s_{\perp}$, and the
external Nambu projectors, without generating an additional matrix
hierarchy.  Solving the resulting recursions gives the complete
frequency-dependent four-point function and the full real-time evolution
from ballistic propagation to Anderson localization.

\begin{figure*}[t]
 \centering
 \includegraphics[width=0.98\textwidth]{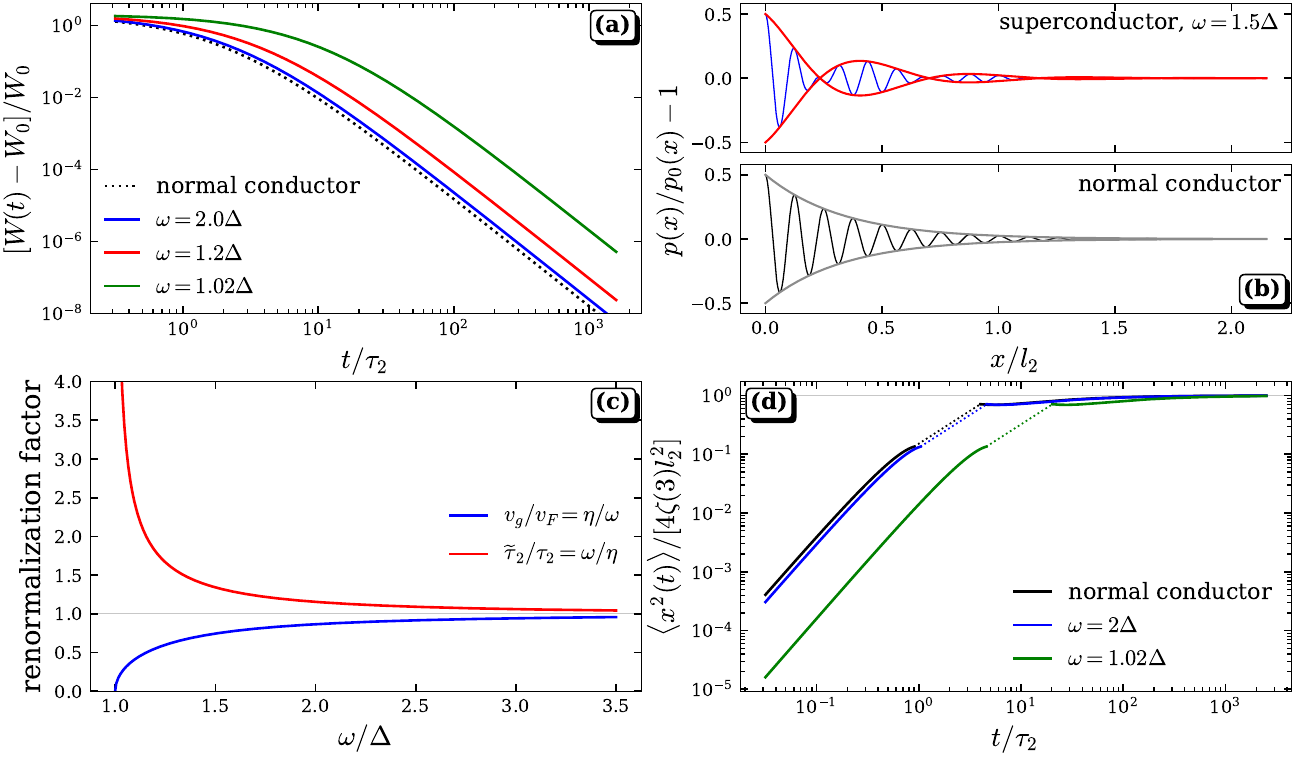}
 \caption{
 Localization dynamics of Bogoliubov quasiparticles.
 (a) Return relaxation,
 $[W(t)-W_{0}]/W_{0}
 =2[2\widetilde{\tau}_{2}/(t+2\widetilde{\tau}_{2})]^{3}$.
 (b) Oscillatory part of the stationary density,
 $p(x)/p_{0}(x)-1
 =\frac{1}{2}e^{-\gamma|x|}
 \cos(2k_{F}x)\cos(2\eta x/v_{F})$,
 with $\gamma=2/l_{1}+1/l_{2}$.
 The normal-state result follows after removing
 $\cos(2\eta x/v_{F})$.
 (c) Reciprocal renormalizations
 $v_{g}/v_{F}=\eta/\omega$ and
 $\widetilde{\tau}_{2}/\tau_{2}=\omega/\eta$, whose product gives
 $v_{g}\widetilde{\tau}_{2}=l_{2}$.
 (d) Mean-square displacement,
 $\langle x^{2}\rangle
 =2(v_{g}t)^{2}[1-2v_{g}t/(3l_{2})]$ at short times and
 $\langle x^{2}\rangle
 =4l_{2}^{2}[\zeta(3)-(2\widetilde{\tau}_{2}/t)
 \ln(t/2\widetilde{\tau}_{2})]$ at long times.
 Dotted lines guide the crossover between the two asymptotic regimes.
 }
 \label{fig:loc_sum}
\end{figure*}

Applying these kernels, we find that the return probability approaches its
stationary value as
$
 W(t)-W_0\propto
 \left(\widetilde{\tau}_2/t\right)^3.
$
The quasiparticle crosses directly from ballistic propagation to
localization, without an intermediate diffusive regime, as shown in
Fig.~\ref{fig:loc_sum}(a).

The mixed-branch sectors describe a specifically superconducting effect.
An injected electron excites both conserved Bogoliubov branches.  Their
momentum difference $2\eta/v_F$ modulates the stationary $2k_F$ density
oscillation by $\cos(2\eta x/v_F)$, producing the long-wavelength
particle--hole beat in Fig.~\ref{fig:loc_sum}(b).  
Its electron--hole interference mechanism is closely related to the geometrical resonances
observed by Tomasch and explained by McMillan and Anderson
\cite{Tomasch1965,McMillan1966}.  Here, the interference appears
in the disorder-averaged localized density profile rather than in
transmission through a finite superconducting film.

The origin of the slow localization dynamics is displayed in
Fig.~\ref{fig:loc_sum}(c).  The group velocity
$v_g=v_F\eta/\omega$ vanishes at the gap edge, while the localization time
$\widetilde{\tau}_2=\tau_2\omega/\eta$ grows by the inverse factor.
Their product remains fixed,
$
 v_g\widetilde{\tau}_2=v_F\tau_2=l_2.
$
Superconductivity therefore delays the formation of the localized state
without changing its spatial scale.

This separation is confirmed by the wave-packet dynamics.  Although
$\widetilde{\tau}_2$ diverges as $\omega\rightarrow\Delta$, the
mean-square displacement saturates at
$
 \langle x^2(t)\rangle=4\zeta(3)l_2^2
$
for every quasiparticle energy, as shown in
Fig.~\ref{fig:loc_sum}(d).  The localized state takes critically longer
to form near the gap edge, but its final extent remains finite and energy
independent.

The unchanged localization length has a complementary spectral
interpretation.  For uniform $\Delta$, each eigenstate $\varphi_n(x)$ of
the disordered normal wire produces a Bogoliubov state with the same
spatial profile and energy
$E_n=\sqrt{\xi_n^2+\Delta^2}$.  Scalar disorder therefore localizes every
above-gap excitation while leaving the average density of states exactly
BCS and generating no subgap states.  This is a dynamical strengthening
of Anderson's theorem for nonmagnetic disorder
\cite{Anderson1959}.  The absence of subgap states relies essentially on
the uniform pair potential in Eq.~\eqref{eq:bdg}.  Interface
inhomogeneity, dissipative broadening, inverse proximity to normal leads,
or pair-breaking disorder can soften the induced gap when this assumption
is relaxed \cite{Takei2013,Stanescu2014}.

Time-resolved thermometry and scanning quasiparticle injection already
resolve propagation and relaxation in superconducting nanowires
\cite{Zgirski2020,Jalabert2023}.  These experiments have not yet entered
the single-channel localized regime considered here, but they provide the
spatial and temporal control needed to test the predicted separation
between localization time and localization length.  
The all-orders kernels also determine observables that are not
fixed by the spectral mapping. Electromagnetic absorption is the
central example because it current vertex introduces Nambu
coherence factors that qualitatively modify the pair-breaking response.

%%%%%%%%%%%%%%%%%%%%%%%%%%%%%%%%%%%%%%%%%%%%%%%%%%%%%%%%%%%%%%%%%%%%%%%

%%%%%%%%%%%%%%%%%%%%%%%%%%%%%%%%%%%%%%%%%%%%%%%%%%%%%%%%%%%%%%%%%%%%%%%
At zero temperature, absorption begins when a photon with
$\nu>2\Delta$ creates two Bogoliubov quasiparticles.  The current operator
is odd under time reversal, so its matrix elements in the normal-state
eigenbasis satisfy $j_{mn}=-j_{nm}$.  The pair-creation amplitude is
$j_{nm}(u_nv_m-v_nu_m)$, and the dissipative conductivity becomes
\begin{equation}
 \sigma_1(\nu)=\frac{\pi e^2}{\nu}
 \int d\xi_1d\xi_2\,
 C(\xi_1-\xi_2){\cal M}(\xi_1,\xi_2)
 \delta(\nu-E_1-E_2),
 \label{eq:absorption-kubo}
\end{equation}
where
$
 {\cal M}(\xi_1,\xi_2)=
 1/2-\left(\xi_1\xi_2+\Delta^2\right)/\left(2E_1E_2\right).
$
This expression separates the superconducting pair-creation factor
${\cal M}$ from the disorder-dependent current kernel $C$, which is the
disorder-averaged density of squared current matrix elements between
normal eigenstates separated by
$\Omega=\xi_1-\xi_2$.

The time-reversal parity of the current produces the minus sign in
${\cal M}$.  Near the pair-breaking threshold,
\begin{equation}
 {\cal M}(\xi_1,\xi_2)
 \simeq\frac{(\xi_1-\xi_2)^2}{4\Delta^2}.
\end{equation}
The current, therefore, cannot create two quasiparticles whose underlying
normal energies coincide.  If $C(\Omega)$ remains finite as
$\Omega\rightarrow0$, this coherence-factor zero produces the linear
Mattis--Bardeen onset \cite{Mattis1958}.

Anderson localization changes the current kernel.  The same kernel
determines the ac conductivity of the normal wire through
$C(\Omega)=\sigma_1^N(|\Omega|)/\pi e^2$.  The all-orders Berezinskii
solution gives
\begin{equation}
 C(\Omega)\simeq\frac{2}{\pi}N(0)l_2^2\tau_2\,
 \Omega^2\ln^2\!\left(2|\Omega|\tau_2\right),
 \qquad |\Omega|\tau_2\ll1 .
 \label{eq:mott-kernel}
\end{equation}
This is the one-dimensional Mott law for resonant absorption between
localized orbitals
\cite{Mott1970,Gogolin1982}.  At small $|\Omega|$, the dominant
contribution comes from rare pairs of orbitals whose energy mismatch is
$|\Omega|$ and whose characteristic separation grows as
$
 l_2\ln\!\left[\frac{1}{|\Omega|\tau_2}\right].
$
In the superconductor, the photon fixes the sum of the Bogoliubov
energies, $E_1+E_2$, while the overlap of the localized orbitals is
controlled by the normal-energy difference $\xi_1-\xi_2$.

This distinction fixes the threshold law.  Let
$\delta\nu=\nu-2\Delta\ll\Delta$.  Energy conservation confines the
normal energies to
$
 \xi_1^2+\xi_2^2=2\Delta\delta\nu.
$
On this threshold circle, the pair-creation factor and the localized
current kernel each vanish as $(\xi_1-\xi_2)^2$.  Performing the angular
integration gives
\begin{equation}
 \sigma_1(\nu)\simeq
 3\pi e^2N(0)l_2^2\tau_2(\nu-2\Delta)^2
 \ln^2\!\left[
 \frac{1}{4\tau_2\sqrt{\Delta(\nu-2\Delta)}}
 \right].
 \label{eq:localized-absorption-edge}
\end{equation}
The excitation threshold remains fixed at $2\Delta$, but Anderson
localization replaces the linear Mattis--Bardeen onset by a quadratic
edge carrying the Mott double logarithm.  It also reduces the
dissipative scale by $(\Delta\tau_2)^2$ relative to the normal Drude
scale. The leading absorption law is therefore reconstructed even
though the single-particle spectrum remains exactly BCS.

\begin{figure}[t]
 \centering
 \includegraphics[width=\columnwidth]{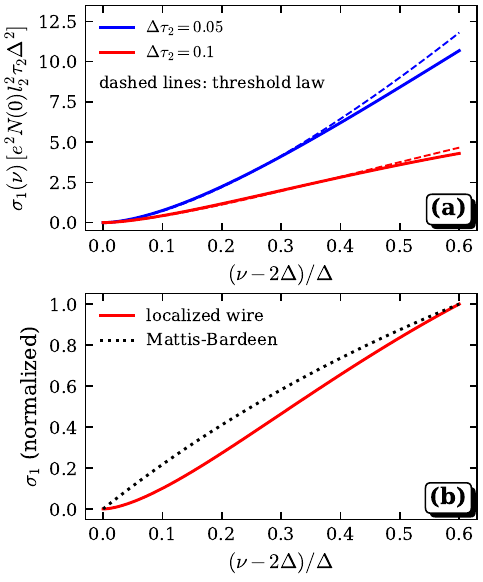}
 \caption{\label{fig:sigma}
 Pair-breaking absorption of the localized wire.
 (a) Full conductivity from Eq.~\eqref{eq:absorption-kubo} (solid) and
 the threshold asymptote in
 Eq.~\eqref{eq:localized-absorption-edge} (dashed).
 (b) Localized and Mattis--Bardeen responses normalized at
 $\nu-2\Delta=0.6\Delta$, exposing their quadratic and linear onsets,
 respectively.
 }
\end{figure}

Fig.~\ref{fig:sigma}(a) shows the full evaluation of
Eq.~\eqref{eq:absorption-kubo} and its approach to the threshold form in
Eq.~\eqref{eq:localized-absorption-edge}.
Fig.~\ref{fig:sigma}(b) compares the resulting quadratic edge with the
linear Mattis--Bardeen onset.  Unlike the anomalous responses observed
near disorder-driven transitions in superconducting films, the effect
obtained here requires neither a soft gap, a spatially fluctuating pair
potential, nor a collective subgap mode.  It follows solely from Mott
correlations between localized quasiparticle orbitals in a hard-gapped
one-dimensional superconductor.

The relevant scales are experimentally accessible.  Taking representative
parameters for epitaxial InAs/Al nanowires,
$\Delta^{*}\simeq0.2\,\mathrm{meV}$,
$v_F\simeq5\times10^{5}\,\mathrm{m/s}$, and
$l_2\simeq0.1$--$1\,\mu\mathrm{m}$, gives
$
 \tau_2=l_2/v_F\simeq0.2\text{--}2\,\mathrm{ps},
 \qquad
 \Delta^{*}\tau_2/\hbar\simeq0.06\text{--}0.6.
$
The corresponding localization length
$L_{\rm loc}=2l_2\simeq0.2$--$2\,\mu\mathrm{m}$ lies within the range of
micron-scale devices provided
$L_{\rm loc}<\min(L,L_\phi)$.  The pair-breaking threshold is
$2\Delta^{*}/h\simeq100\,\mathrm{GHz}$, and the Mott asymptote applies
when
$
 4\tau_2\sqrt{\Delta\,\delta\nu}\ll1,
 \qquad
 \delta\nu=\nu-2\Delta,
$
or equivalently
$\delta\nu\ll(16\Delta\tau_2^2)^{-1}$.  Broadband microwave or
millimeter-wave spectroscopy can therefore distinguish the predicted hard
threshold, quadratic onset, and reduced dissipative weight from the
Mattis--Bardeen response.

At finite temperature, subgap photons probe pre-existing quasiparticles
rather than creating a pair, a regime already accessed in superconducting
microwave resonators \cite{deVisser2014}.  In equilibrium linear response,
photons with $\nu<2\Delta$ scatter thermally excited quasiparticles within
the same branch.  The scattering coherence factor remains finite near the
gap, while the photon frequency is converted into the larger normal-energy
separation $\nu E/\sqrt{E^2-\Delta^2}$.  For
$\nu\ll T\ll\Delta$, we obtain
\begin{equation}
 \sigma_1^{\rm th}(\nu)\simeq
 e^{-\Delta/T}\,
 \sigma_1^N\!\left(
 \nu\sqrt{\frac{\pi\Delta}{2T}}
 \right).
 \label{eq:thermal-absorption}
\end{equation}
Thermal quasiparticles, therefore, retain the localized absorption kernel
of the normal wire, but their density is activated, and the frequency
entering that kernel is stretched by superconductivity.

\begin{figure}[t]
 \centering
 \includegraphics[width=\columnwidth]{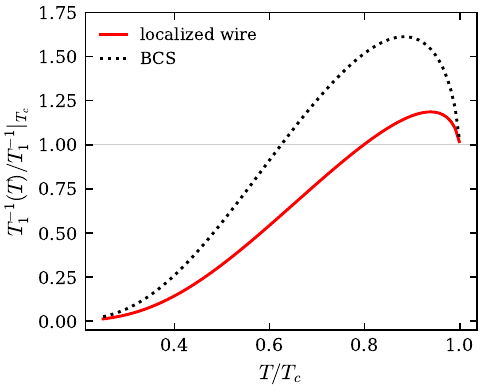}
 \caption{\label{fig:HS}
 Normalized nuclear spin-relaxation rate from
 Eq.~\eqref{eq:T1-ratio}.  The dotted curve is the broadened BCS result
 with $\Lambda=1$.  The solid curve includes localization with
 $\tau_{\rm in}/(2\tau_2)=3$ and
 $\omega_e\tau_{\rm in}=1$.  Both curves use the same Dynes width
 $\Gamma=0.04\Delta_0$, where $\Delta_0=\Delta(T=0)$.
 }
\end{figure}

The same spectral stretching modifies the Hebel--Slichter coherence peak
in nuclear spin relaxation \cite{HebelSlichter1959}.  A nuclear spin flip
scatters an existing quasiparticle with a coherence factor
$(u_1u_2+v_1v_2)^2$, which remains finite at the gap edge.  Defining
$
 \eta_E=\sqrt{E^2-\Delta^2},
 \qquad
 N_s=E/\eta_E,
 \qquad
 D_s=\Delta/\eta_E,
$
we obtain
\begin{align}
 \frac{T_1^{-1}(T)}{T_1^{-1}|_{T_c}}
 &=2\int_\Delta^\infty dE
 \left(-\frac{\partial f}{\partial E}\right)
 \bigl[N_s^2(E)+D_s^2(E)\bigr]\Lambda(E),
 \label{eq:T1-ratio}
\end{align}
where
\begin{eqnarray}
 \Lambda(E)
 &=&
 \left[
 1+\frac{\tau_{\rm in}}{2\tau_2}
 \frac{1}{1+(\omega_e\tau_{\rm in}E/\eta_E)^2}
 \right]
 \nonumber\\
 &&\times
 \left[
 1+\frac{\tau_{\rm in}}{2\tau_2}
 \frac{1}{1+\omega_e^2\tau_{\rm in}^2}
 \right]^{-1}.
\end{eqnarray}
The factor $\Lambda(E)$ describes repeated returns of a localized
quasiparticle to the nuclear site, with the return singularity broadened
by $\tau_{\rm in}$.  A nuclear energy transfer $\omega_e$ corresponds to
the larger normal-energy separation $\omega_e E/\eta_E$.  This separation
diverges as $E\rightarrow\Delta$ and removes resonant localized pairs from
the nuclear-frequency window.  The localization enhancement is therefore
switched off precisely where the BCS factor $N_s^2+D_s^2$ is largest.
The resulting reduction of the Hebel--Slichter peak is shown in
Fig.~\ref{fig:HS}.  Localization thus provides a mechanism for suppressing
the coherence peak without changing the pairing symmetry or introducing
pair-breaking disorder.

In summary, by extending the Berezinskii resummation to a single-channel
superconducting wire, we have obtained its disorder-averaged quasiparticle
dynamics to all orders in elastic scattering.  Scalar disorder preserves
the hard BCS spectrum and the normal-state localization length, but
separates the temporal and spatial scales of localization.  Near the gap
edge, the time required to form a localized state diverges while its
final spatial extent remains finite.

Electromagnetic absorption reveals a distinct consequence of localization
that is not captured by the unchanged BCS spectrum.  The pair-creation coherence factor and Mott correlations
between localized orbitals combine to convert the linear
Mattis--Bardeen onset into a parametrically weaker quadratic edge with a
double logarithm.  The same spectral stretching governs thermal
absorption and reduces the Hebel--Slichter peak.  These results establish
an exact distinction between spectral and dynamical consequences of
disorder.  Anderson localization can leave the superconducting spectrum
unchanged while reconstructing the dissipative response above the gap.
%%%%%%%%%%%%%%%%%%%%%%%%%%%%%%%%%%%%%%%%%%%%%%%%%%%%%%%%%%%%%%%%%%%%%%%

%%%%%%%%%%%%%%%%%%%%%%%%%%%%%%%%%%%%%%%%%%%%%%%%%%%%%%%%%%%%%%%%%%%%%%%
%{\it Acknowledgments.}
This work is supported in part by the Turkish Academy of Sciences 
(TUBA) under Grant No. AD-2026.
%%%%%%%%%%%%%%%%%%%%%%%%%%%%%%%%%%%%%%%%%%%%%%%%%%%%%%%%%%%%%%%%%%%%%%%

%\nocite{*} 
\newpage
\bibliography{paper}% Produces the bibliography via BibTeX.

\end{document}